\documentclass[12pt]{article}
\usepackage{amsfonts}
\usepackage{amsmath}
\usepackage{setspace}
\usepackage{graphicx}
\usepackage{psfrag}
\usepackage{graphics}

\begin{document}
\title{On the Regularizability of the Big Bang Singularity
} 

\date{September 17, 2012}
\maketitle
\centerline{\scshape Edward Belbruno} 
\medskip
{\footnotesize
    \centerline{Princeton
University, Department of Astrophysical Sciences}
\centerline{belbruno@princeton.edu}  
}

\medskip\medskip\medskip

\bigskip

\maketitle

\begin{abstract}
The singularity for the big bang state can be represented using the generalized anisotropic Friedmann equation, resulting in a system of differential equations in a central force field. We study the regularizability of this singularity as a function of a parameter, the equation of state, $w$. We prove that for $w >1$ it is regularizable only for $w$ satisfying relative prime number conditions, and for $w \leq 1$ it can always be regularized. This is done by using a McGehee transformation, usually applied in the three and four-body problems. This transformation blows up the singularity into an invariant manifold. The relationship of this result to other cosmological models is briefly discussed. 
\end{abstract}

\section{\bf{Introduction}}
\label{intro}

It is well known that the Kepler problem, which defines the motion of two bodies moving in a Newtonian gravitational inverse square central force field, can be regularized at the collision of the two bodies. Letting $r$ represent the distance between the two bodies, then the regularization of collision implies that there exists a change of position and velocity coordinates, and time, so that in the new coordinates and time the state corresponding to $r=0$ is well defined. Moreover, the flow of the transformed differential equations in position and velocity space is smooth(real analytic) in a neighborhood of collision, and also on the entire associated fixed energy surface. This implies that solutions can be smoothly extended across collision as a function of time. The Levi-Civita transformation is a classical example of a regularization that accomplishes this in the two-dimensional Kepler problem (Levi-Civita 1920). Physically, this means that the two bodies perform a smooth bounce at collision.  Within the field of celestial mechanics, there are many other regularizations for binary collisions in the Kepler problem. These include the use of geodesic flow equivalent transformations, due to Moser and this author (Moser 1970; Belbruno 1977), and the Kustaanheimo-Stiefel transformation (Kustaanheimo and Stiefel 1965).  

Regularization transformations can be applied to problems other than the classical Newtonian inverse square gravitational force, to which they have been generally historically focused. The regularization of the collision of a zero mass particle, eg a photon,  with a Schwarzschild black hole was described in a recent paper by F. Pretorius and this author (Belbruno and Pretorius 2011). This is accomplished by equivalently viewing the motion of the particle as being in an inverse forth power central force. The differential equations describing this motion can be transformed using a mapping due to R. McGehee (McGehee 1981), which blows up the collision state into an invariant manifold. This transformation describes the motion of the particle about the black hole is a way that eludicates the entire phase space in a surprisingly clear fashion. At the same time, the collision state of the particle with the black hole can be regularized.  This turns out to be a 'branch regularization', where individual solutions have unique extensions across collision as a function of time, which, in position coordinates, are only continuous at collision and not differentiable there. The key point is that the extension across collision is {\em uniquely} defined.  This implies that when the particle collides with the black hole, it performs a continuous, but not differentiable, bounce to a unique extension, or bounce, from the black hole.

In this paper we can gain interesting information about the dynamics in a neighborhood of the singularity due to the big bang using a McGehee transformation, following the same approach used in Belbruno and Pretorius(2011). We first show that dynamical flow near the big bang singularity can be reduced to a central force field, when modeled by an anisotropic Friedmann equation, under a number of assumptions.  This force field is defined with inverse powers of $a$, $a^{-\kappa}$, $\kappa \geq 5$,  where $a$ is the expansion factor of the universe and $a=0$ corresponds to the big bang. We then apply the McGehee transformation to the central force field, yielding unique branch extensions of solutions through $a=0$. 

We prove that the big bang can be branch regularized, provided that one of the parameters, the equation of state variable, $w > 1$, satisfies relative prime number conditions, which are a set of measure zero.  In other words, for $w > 1$, except for a set of measure zero, the big bang is not branch regularizable. Under these assumptions, this implies that if we consider a universe prior to ours, for $t >0$ that collapsed to $a=0$ at time $t=0$, then as a function of time, our universe for $a(t) > 0$, with $t < 0$, cannot be defined as a branch extension, from the previous universe for $w > 1$, except for a set of measure zero. However, for $w \leq 1$, the singularity can always be branch regularized. The equation of state, $w$, is a key parameter in cosmology. $w = P/\rho$, where $P$ is the pressure and $\rho$ is the energy density of the universe. The parameter $w$ and relevance of $w=1$ is explained further in Section ~\ref{sec:3}. It is noted that we have chosen the convention that $t > 0$ prior to the big bang singularity and $t < 0$ after the big bang. This is chosen for convenience to be consistent with the time convention used in McGehee(1981), since McGehee(1981) is referred to frequently. 

This result can be related to other cosmological models, which we do in Section ~\ref{sec:3}.  It is important to note that our results are purely mathematical in nature and we are not making any assertions to the relevance of these results to reality.

The paper is organized as follows: In Section ~\ref{sec:1} the problem is formulated. Section ~\ref{sec:2} gives the solution to the problem and the main results. Block regularization is discussed at the end of Section ~\ref{sec:2}. In the final section, Section ~\ref{sec:3}, we briefly discuss the relevance of our results to other cosmological models.

\section{\bf{Formulation of Problem}}
\label{sec:1} 

To formulate the problem, we begin with the {\it cosmological principle}. This states that at any moment in time, the universe presents the same aspect from every point, except for local irregularities. This equivalently means that the universe is isotropic and homogeneous.  

Assume that we know the distribution of the total finite number of galaxies at time $t_0$ of distance $r_i$ from a fixed origin $O$, $i=1,..., N$ where $N$ is the total number of galaxies. Then the cosmological principle implies that $r_i(t) = a(t)r_i(t_0)$, where $t\in \Re$ is time. $a(t)$ is called the Friedmann-Robertson-Walker(FRW) scale factor. That is, the cosmological principle implies a uniform expansion or contraction. Hubble's law implies,
\begin{equation}
H^2 = \frac{\dot{a}^2}{a^2}, 
\label{eq:Hubble}
\end{equation}
where $H = H(t)$ is the Hubble parameter, $a = a(t)$, $^. \equiv d/dt$. $H(t_0)$ is the Hubble constant, which is the current expansion rate of the universe.

We consider the Friedmann equation that approximates cosmic evolution,   
\begin{equation}
H^2 = \frac{8\pi G}{3}  \left(  \frac{\rho^0_m}{a^3} + \frac{\rho^0_\tau}{a^4} + \frac{\rho^0_w}{a^{3(1+w)}} \right) - \frac{K}{a^2} + \frac{\sigma^2}{a^6},
\label{eq:H2}
\end{equation} 
where $a=a(t)$ is normalized, so that for today($t=t_0 < 0$) $a(t_0)=1$. 
$\rho^0_j$ is the present value of the energy density for component $j$, where $m$ represents non-relativistic matter, $\tau$ represents
radiation and $w$ represents an energy component with equation of state $w$. 
The last two terms on the right-hand side represent the spatial curvature and anisotropy, respectively. $K$ is the curvature and $\sigma$ is
the anisotropy parameter(see Garfinkle et al. 2008). 
\medskip

We are interested in studying solutions where $a \rightarrow 0$.   This is initially  assumed to take
place in a contracting phase of the universe, just prior to the big bang, where $t>0$. 
This assumption is consistent with current alternative theories in cosmology to the
standard big bang inflationary model, such as the cyclic or ekpyrotic models (Steinhardt and Turok 2002; Garfinkle et al. 2008), where the
universe undergoes a period of slow contraction with $w > 1$ prior to the big bang. 
We are assuming that the big contraction(crunch) converges to $a = 0$. 

The goal  of this 
paper is to understand under what conditions solutions $a=a(t)$ can be continued, in a well defined manner, with $t > 0$, as a function of time 
to $t=0$,  which corresponds to the start of the big bang, and then for  $t < 0$ beyond the big bang. The way this will be done is to see under
which conditions, the function  $a(t)$ can be regularized at $t=0$.  The type of regularization we are using is branch regularization, mentioned in the Introduction,  where the trajectory
for $a(t)$ for $t > 0$ is smooth(analytic) and can be uniquely extended to  $t \leq  0$, which is continuous, at  $t=0$ and analytic for $t < 0$ (see McGehee 1981).
Branch regularization yields a continuous well defined unique bounce at $t=0$ for the given trajectory with initial condition in the contracting phase for $t>0$. 
We explain how this can be accomplished in the next section, where branch regularization is carefully defined. 
\medskip

\section{{\bf Assumptions, Differential Equations, and  Results}} 
\label{sec:2}

Friedmann's equation, (\ref{eq:H2}), gives rise to a second order differential equation for $a(t)$ under some assumptions on the cosmology model we are using. 
\medskip\medskip

\noindent
{\em Assumption A} \hspace{.1in}  Let $t= t_1$ be the initial time for $a(t)$ in the contracting phase of the universe prior to the big bang at $t = 0$, so
that $0 < t_1 \ll 1$.  Since it is assumed that $a \rightarrow 0$ as $t \rightarrow 0^+$ ($0^+$ means $t \rightarrow 0$ with $t >0$), 
then $a(t_1)$ is assumed to be sufficiently small, say $0 < a(t_1) < \delta \ll 1 $. 
\medskip

\noindent
{\em Assumption B}  \hspace{.1in} $w$ and $K$ are constant. $\sigma, \rho^0_m, \rho^0_\tau, \rho^0_w$ are positive constants.    
\medskip\medskip   

\noindent
It is reasonable to fix these parameters in a neighborhood of $a=0$. For example, in the ekyprotic model of the big bang, where $w \geq 1$, numerical computations show that these parameters generally converge to constant values (Garfinkle et al. 2008).   
\medskip  

Under Assumption B, we differentiate (\ref{eq:Hubble}) with respect to t. We first substitute (\ref{eq:H2}) into(\ref{eq:Hubble}), which yields
\begin{equation}
{\dot{a}}^2 - \frac{\sigma^2}{a^4} + K - \frac{8\pi G}{3}  \left(  \frac{\rho^0_m}{a} + \frac{\rho^0_\tau}{a^2} + \frac{\rho^0_w}{a^{3(1+w)-2}} \right) = 0.
\label{eq:LargeEqu} 
\end{equation}
Applying $d/dt$ to both sides of (\ref{eq:LargeEqu}) yields
\begin{equation}
\ddot{a}\dot{a} + \frac{2\sigma^2}{a^5} \dot{a} + \frac{4\pi G}{3} \dot{a}\left( \frac{\rho^0_m}{a^2} + \frac{2\rho^0_\tau}{a^3} 
+ \frac{\rho^0_w [3(1+w)-2]}{a^{3(1+w)-1}} \right) = 0.
\label{eq:LargeEqu2}
\end{equation}
\noindent
We can divide through both sides of (\ref{eq:LargeEqu2}) by $\dot{a}$ which is nonzero. This follows from (\ref{eq:LargeEqu}) which implies that 
as $a \rightarrow 0$, $ |\dot{a}| \rightarrow \infty$.   
After rearranging terms, we obtain,
\begin{equation}
\ddot{a} = -  \frac{2\sigma^2}{a^5} -  \frac{4\pi G}{3} \left( \frac{\rho^0_m}{a^2} + \frac{2\rho^0_\tau}{a^3}
+ \frac{\rho^0_w [3(1+w)-2]}{a^{3(1+w)-1}} \right) 
\label{eq:LargeDE}
\end{equation}

The differential equation given by (\ref{eq:LargeDE}) for $a=a(t)$ plays a key role in our analysis. 
It defines the behavior of $a(t)$ as a function of $w$. We will consider three cases: $w < 1$, $w = 1$, and $w > 1$. 
\medskip
\medskip

The term 'branch regularization', mentioned in the Introduction, is formally defined as in McGehee(1981). We define it for a general first order system of differential equations,
\begin{equation}
\bf{\dot{x} = F(x)}
\label{eq:Gen}
\end{equation}  
where $^. \equiv d/dt$, ${\bf x} \in R^n$, $n \geq 2$. Assume $\bf{F}$ is a real analytic vector field on an open set $U \subset R^n$. 
\medskip

We first define the term 'singularity',
\medskip

\noindent
Definition 1 \hspace{.1in} Let ${\bf x}(t)$ be a solution for (\ref{eq:Gen}), defined in $U$ for $b < t < a$, with initial condition ${\bf x}(t_0)$,  $ -\infty \leq b < t_0 < a \leq \infty$. $(b,a)$ is assumed to be the maximal interval the solution can be extended.  If $ -\infty < b$ then the solution is said to begin at $t = b$ and if $a < \infty$, then the solution is said to end at $t = a$. In either case, $t^* = a$ or $t^* = b$ is said to be a {\em singularity} for ${\bf x}(t)$. 
\medskip

The terms 'branch extension' and 'branch regularization' are defined in Definitions 2,3, respectively,
\medskip

\noindent
Definition 2  \hspace{.1in} Let ${\bf X_1}(t) , {\bf X_2}(t)$  be two solutions of (\ref{eq:Gen}). Suppose that $\bf{X_1}$ ends in a singularity at $t^*$ and $\bf{X_2}$ begins in the singularity at the same time. Assume there is a multivalued analytic function having a branch at $t^*$, and extending both ${\bf X_1},{\bf X_2}$, where we regard $t$ as complex.  Then $\bf{X_1}$ is said to be a {\em branch extension} of $\bf{X_2}$ and $\bf{X_2}$ is said to be branch extension of $\bf{X_1}$.
\medskip

\noindent
Definition 3 \hspace{.1in} A solution $\bf{X}$ of (\ref{eq:Gen}) which either begins or ends in a singularity at time $t^*$
is called {\em branch regularizable} at $t^*$ if it has a unique branch extension at $t^*$.
\medskip

It is important to note that in Definitions 2,3, time $t$ is assumed to be complex. The solutions $\bf{X_1}, \bf{X_2}$ are real analytic, taking on real values for real $t$. However, they are extensions of each other through complex $t$. This definiton of branch regularization is well defined for one degree of freedom, n=1, as is seen, using time as a complex variable.  
\medskip

We are interested in obtaining branch extensions for (\ref{eq:LargeDE}) at $t = t^* = 0$. (\ref{eq:LargeDE}) is written as a first order system analogous to (\ref{eq:Gen}) by setting ${\bf x} = (a, \dot{a})$. We need only find a branch extension for  
for $a(t)$ since that automatically yields one for $\dot{a}(t)$.  
\medskip

Branch regularization is defined along trajectories of solutions. It's main feature is that it yields unique extensions at singular
points whose solutions are not defined in the original coordinates. Branch regularization at $a=0$ for (\ref{eq:LargeDE}) will be proven by a change of coordinates
and time, $a, \dot{a}, t$, respectively, with an explicit transformation, to new coordinates, $r, v, s$, respectively, where the state $a=0, |\dot{a}| = \infty$ 
is well defined, corresponding to $r=0, v = \pm\sqrt{2}$. 
In this way, the singular state at $a=0$ is resolved.
The solution, $a(t)$,  is then well defined and continuous at $t=0$. This is obtained through an explicit formula.   
\medskip

\noindent
It is remarked that although McGehee's work (McGehee 1981) is formulated in general
for two degrees of freedom, it can be reduced to one degree of freedom, as we have applied in this paper, in a straight forward manner, when extending a solution $a(t)$ from $t >0$ to $t \leq 0$.  The proof of a unique extension in McGehee(1981), in general,  requires knowing the explicit dependence of the solution, in this case $a(t)$, as a function of $t$ near $t=0$ by transforming $a$ and $\dot{a}$ to regularized coordinates, $r,v$, using the original time coordinate $t$. In the regularized coordinates, the explicit form of the solution $r(t)$ can be obtained using the stable manifold theorem, where uniqueness is established using both $r,v$. This yields a unique branch in these coordinates. Transforming from $r$ to $a$ yields a unique branch for $a(t)$. This is described in more detail in the proof of Theorem 5. 
\medskip

We are now able to state our results. 
\medskip

\noindent
{\bf Theorem 1 }  
\medskip

\noindent
(a.)\hspace{.1in} If $|w| \leq 1$, then every solution $a \rightarrow 0$ can be branch regularized at $a=0$. 
\medskip

\noindent
(b.)\hspace{.1in} If $w >1$, then a branch regularization of $a=0$ can be done on a countable set (measure zero), $Q_w$, of $w$ values, 
or equivalently, a branch regularization of $a=0$ does not exist for almost all values of $w$ (i.e. outside the set $Q_w$). The set $Q_w$ is given by\\

\centerline{  $Q _w = \{ w = \frac{2}{3} \frac{q}{p} - 1\hspace{.1in} | (p,q) \in \wp     \}$,       }
\medskip\medskip

\noindent 
where $p, q$ are positive integers and 
\medskip\medskip

\centerline{ $ \wp = \{ (p,q) | 0 < p < q , \hspace{.1in} (p,q) \hspace{.1in}$ relatively prime, $ \hspace{.1in} q\hspace{.1in}$  odd \}.}  
\medskip
\medskip

\noindent 
The condition of regularizability with $w \in Q_w$ is both necessary and sufficient.
\medskip
\medskip
\medskip
\medskip

We now show how this theorem is proven in a series of theorems.
\medskip\medskip

\noindent A preliminary constant time scaling is done from $t$ to $\tau$ (this use of $\tau$ is not to be 
confused with its earlier use in the symbol, $\rho^0_{\tau}$),    
\begin{equation}
t = \begin{cases}  2 (2)^{-1/2}\sigma^{-1}  \tau, &\mbox{if} \hspace{.1in} w < 1,\\
2C^{-1/2} \tau, &\mbox{if} \hspace{.1in} w = 1, \\
((4/3)\pi G \rho^0_w)^{-1/2} \tau, &\mbox{if} \hspace{.1in} w > 1, 
\end{cases}
\label{eq:TimeScale}
\end{equation}  
where $$C = 2\sigma^2 + (16/3)\pi G \rho^0_w.$$ By Assumption B, this time scaling is well defined, where $C > 0$.

\noindent
As $a \rightarrow 0$, we can take $a$ sufficiently small so that the term with the largest power of $a^{-1}$ in the right hand side 
of (\ref{eq:LargeDE}) dominates the behavior of $\ddot{a}$. In fact, the following theorem 
is directly obtained,   
\medskip\medskip\medskip

\noindent
{\bf Theorem 2} \hspace{.1in}  In the scaled time variable, $\tau$, given by (\ref{eq:TimeScale}), 
and where $' \equiv d/d\tau$, the differential equation (\ref{eq:LargeDE}), can be reduced to the following:
\medskip

\noindent
If $w < 1$, 
\begin{equation}
a'' = -\frac{4}{a^5}  -  f_1(a),
\label{eq:SmallDE1}
\end{equation}
where  
\begin{equation}
f_1(a) =  a_1 a^{-2} + a_2 a^{-3} + a_3 a^{-(5-\kappa_1)},
\label{eq:f1}
\end{equation}
$\kappa_1(w)= 3(1-w) >0$ is a constant, $a_k$ are constants, and $\kappa_1 \rightarrow 0^+$ as $w \uparrow 1$. If we set $\tilde{c} = (4/3)\pi G$, then $a_1 = 2\sigma^{-2}{\tilde{c}}\rho^0_m$, $a_2 = 4\sigma^{-2}{\tilde{c}}\rho^0_{\tau}$, $a_3 = 2 \sigma^{-2} {\tilde{c}}(4-\kappa_1) \rho^0_w. $  Thus the values of the powers of $a^{-1}$ of $f_1$ are less than that of the first term on the right side of (\ref{eq:SmallDE1}). 
\medskip\medskip

\noindent
If $w=1$
\begin{equation}
a'' = -\frac{4}{a^5} - f_2(a),
\label{eq:SmallDE1=1}
\end{equation}
where
\begin{equation}
f_2(a) = b_1 a^{-2} + b_2 a^{-3}, 
\label{eq:f2}
\end{equation}
and $b_k$ are constants; $b_1 = 4\tilde{c} C^{-1} \rho_m^0$, $b_2 = 8\tilde{c} C^{-1} \rho_{\tau}^0$.  
\medskip\medskip

\noindent
If $w > 1$,
\begin{equation}
a'' = - \frac{4 + \kappa_2}{a^{5+\kappa_2}}  - f_3(a), 
\label{eq:SmallDE2}
\end{equation}
where
\begin{equation}
f_3(a)  = c_1 a^{-2} + c_2 a^{-3} + c_3 a^{-5},
\label{eq:f3}
\end{equation}
where $\kappa_2 = 3(w-1) > 0$ is a constant, $c_k$ are constants, and $\kappa_2 \rightarrow 0^+$ as $w \downarrow 1$; $c_1 = \rho^0_m {\rho^0_w}^{-1}$,$c_2 = 2 \rho^0_{\tau} {\rho^0_w}^{-1}$, $c_3 = 2\sigma^2 {\rho^0_w}^{-1}$.  
Thus, the values of the powers of $a^{-1}$ of $f_3$ are less than that of the first term on the right side of (\ref{eq:SmallDE2}). 
\medskip

The coefficients $a_k, b_k, c_k$ are well defined by Assumption B.
\medskip\medskip\medskip\medskip\medskip

The differential equations, (\ref{eq:SmallDE1}), (\ref{eq:SmallDE1=1})  (\ref{eq:SmallDE2}), can be written as
\begin{equation}
a'' = \frac{dV}{da},
\label{eq:DEPot}
\end{equation}
where $V$ is a central force potential function,
\begin{equation}
V = \frac{1}{a^{\alpha}} + f(a), 
\label{eq:Pot}
\end{equation}
where 
\medskip
$$
\alpha = \begin{cases} 4, &\mbox{if} \hspace{.1in}  w \leq 1, \\
3(1+w)-2, &\mbox{if} \hspace{.1in} w > 1.  \end{cases}
$$
\noindent
$f(a) = O(a^{-y}), y < \alpha$ is a sum of terms of the form $a^{-\rho_k}$, 
$\rho_k < \alpha$.   
\medskip

\noindent
More exactly, 
$$
f(a) = \begin{cases}  a_1 a^{-1} + \frac{1}{2} a_2a^{-2} + \frac{a_3}{4-\kappa_1} a^{-(4-\kappa_1)} , &\mbox{if} \hspace{.1in} w < 1, \\
b_1 a^{-1} + \frac{1}{2} b_2 a^{-2}, &\mbox{if} \hspace{.1in}  w = 1, \\
c_1 a^{-1} + \frac{1}{2} c_2 a^{-2} + \frac{1}{4} c_3 a^{-4}, &\mbox{if} \hspace{.1in} w > 1, 
\end{cases}
$$
where it is noted that the term $a_3/(4-\kappa_1)$ is well defined for $\kappa_1 = 4$.     
\medskip\medskip\medskip

\noindent
The new variables are now $a, P = a', \tau$ and the new differential equation is (\ref{eq:DEPot}). 
This can equivalently be written as a first order Hamiltonian system of one degree of freedom, 
\begin{equation}
a' = \frac{\partial H}{\partial P}, \hspace{.2in} P ' = - \frac{\partial H}{\partial a},
\label{eq:HamSys}
\end{equation}
where,
\begin{equation}
H = \frac{1}{2}P^2   -  V(a).
\label{eq:Hamiltonian}
\end{equation}
\medskip
\noindent
$H$ is an integral of the motion, so that each trajectory $X(\tau) = (a(\tau), P(\tau))$ lies on the energy manifold
\begin{equation}
\Sigma = \{ a,P| H(a,P)=h \},
\label{eq:EnergySurface}
\end{equation}
for some constant $h \in R^1$. $\Sigma$ is a one-dimensional curve. $h$ is uniquely determined from the initial value of $X(\tau)$. It is 
seen that on each manifold, $\Sigma$, $|P| \rightarrow \infty$ as  $a \rightarrow 0$. 
\medskip\medskip

\noindent
It is noted that by the time scaling (\ref{eq:TimeScale}), multiplying $t$ by a positive constant, $A$, $\tau =A t$, the results we obtain wrt $\tau$ are immediately obtained for $t$, where
$\tau \rightarrow 0$ is equivalent to $t \rightarrow 0$.
Thus, we will present the results from this point on for (\ref{eq:HamSys}) without loss of generality.
\medskip\medskip

\noindent
Set  $\gamma = (1 + \beta)^{-1}$ and $\beta = \alpha /2$, 
We apply a transformation that smooths the singularity at $a=0$ so that $|P| = \infty$ gets mapped into a well defined set. 
\medskip\medskip\medskip

\noindent
{\bf Theorem 3} \hspace{.1in} The transformation $T_M: a, P \rightarrow r, v$,  
of (\ref{eq:DEPot}), or equivalently (\ref{eq:HamSys}), which smooths the singularity at $a=0$, is given by
\medskip
\begin{equation}
a = r^{\gamma},  \hspace{.2in} 
P = r^{-\beta \gamma}v, 
\label{eq:Trans}
\end{equation}
where, if $w \leq 1$,  
\begin{equation}
\alpha = 4, \hspace{.2in} \beta = 2, \hspace{.2in}  \gamma = 1/3 ,
\label{eq:Values1}
\end{equation}
and if $w > 1$,
\begin{equation}
\alpha  = 3(1+w)-2 , \hspace{.2in} \beta = \frac{3}{2}(1+w)-1, \hspace{.2in} \gamma = (1 + \beta)^{-1}.  
\label{eq:Values2}
\end{equation}
This case implies that $0 < \gamma < 1/3$ and $\beta > 2$. 
\medskip

\noindent
The Hamiltonian energy (\ref{eq:Hamiltonian}) gets mapped into the transformed energy equation,
\begin{equation}
v^2 - 2 = 2hr^{\alpha\gamma} + 2r^{\alpha \gamma} g(r),
\label{eq:TransEnergy}
\end{equation}
where 
$$ g(r) = f(r^{\gamma}) .$$
\medskip\medskip\medskip\medskip

\noindent The proof of this is obtained by noting that the transformation $T_M$ is directly obtained from a more general map for two-degrees of freedom in McGehee(1981)(see Equation 4.1, where $w =0, \theta = 0$, and Equation 4.7). It is similarly verified that (\ref{eq:TransEnergy}) is obtained(see Equation 4.3 in McGehee(1981)(see also Belbruno and Pretorius 2011)). In our case, we have the additional term $f(a)$ which gives rise to the term $2r^{\alpha\gamma} g(r) = O(r^{(\alpha - y)\gamma})$ in (\ref{eq:TransEnergy}),  where $\alpha - y > 0$, and $\gamma > 0$.  More exactly,
\begin{equation}
\tilde{G}(r) = 2r^{\alpha\gamma} g(r) = \begin{cases}  a_1 r^{} + \frac{1}{2} a_2r^{2/3} + \frac{a_3}{4-\kappa_1} r^{\kappa_1/3} , &\mbox{if} \hspace{.1in} w < 1, \\
b_1 r^{} + \frac{1}{2} b_2 r^{2/3}, &\mbox{if} \hspace{.1in}  w = 1, \\
c_1 r^{\gamma(3+\kappa_2)} + \frac{1}{2} c_2 r^{\gamma(2+\kappa_2)} + \frac{1}{4} c_3 r^{\gamma\kappa_2}, &\mbox{if} \hspace{.1in} w > 1. 
\end{cases}
\label{eq:EnergyTerm}
\end{equation}

\noindent
Thus, the energy manifold $\Sigma$ is  mapped into
\begin{equation}
M = \{ r,v |  v^2 - 2 = 2hr^{\alpha\gamma} + \tilde{G}(r) \}  
\label{eq:EneryManifold}
\end{equation}
Since $a=0$ is mapped into $r=0$, and $\tilde{G}(0) = 0$, then this defines the 'collision
manifold' $N$ (see McGehee 1981), which represents the big bang singularity,
$$ N = \{ (r,v) \in M | r=0, v=\pm\sqrt{2} \}. $$ Thus, $|P| = \infty$ is mapped into $v=\pm\sqrt{2}$. 
A set analogous to this is defined in Belbruno and Pretorius(2011) for the problem of
motion about a black hole, using the two-degree of freedom version of $T_M$. $N$ is an invariant manifold for the transformed flow in a new time variable we define below.   
\medskip\medskip

\noindent 
The transformation $T_M$ smooths the singularity at $a=0$ in the sense that $|P|=\infty$ is mapped into a finite set of points. 
To see if a solution $a=a(\tau)$ can be extended through collision as a function of $\tau$, it is 
necessary to examine the transformed differential equations. 
\medskip\medskip

\noindent
Set 
\begin{equation}
Q_{\gamma} = \{   \gamma  = \frac{p}{q} \hspace{.1in} | (p,q) \in \wp  .    \}
\label{eq:GammaCondition}
\end{equation}
Adapting a theorem in McGehee(1981) of two degrees of freedom to one degree of freedom, and in particular to our system (\ref{eq:HamSys}), and also using Equation (\ref{eq:Solution}), proven to exist below in the proof of Theorem 5,  we have
\medskip\medskip\medskip

\noindent
{\bf Theorem 4 (McGehee)}  \hspace{.1in} Every singular solution $a \rightarrow 0$ to (\ref{eq:HamSys}) can be
branch regularized  if and only if $\gamma \in  Q_{\gamma}$.
\medskip\medskip\medskip

\noindent
This implies that if $\gamma \notin Q_{\gamma}$, then no singular solution can be branch regularized, and if $\gamma \in Q_{\gamma}$, then
every singular solution can be branch regularized.
\medskip

\noindent
The relative prime number condition in Theorem 4 is a key condition for this analysis. It results from a more general theorem in McGehee(1981), proven for the more general case of two-degrees of freedom for (\ref{eq:HamSys}),
where, more generally,
$a=(a_1, a_1)$, $P=(P_1, P_2)$, and in (\ref{eq:Hamiltonian}), $P^2 \equiv |P|^2$, $a^{\alpha} \equiv |a|^{\alpha}$. However, it can be applied to our case
of a single degree of freedom in a straight forward manner, as is verified.
\medskip

\noindent
Before we look at the transformed flow and the explicit form of the solution $a(\tau)$, we discuss the set $Q_{\gamma}$. 
\medskip
\medskip

\noindent
The prime number condition that $\gamma \in  Q_{\gamma}$ is equivalent to a prime number condition on $\alpha$ and $w$. This follows from the
definition of $\gamma = (1 + (\alpha/2))^{-1}$ which implies,
$$\alpha = 2(\gamma^{-1} - 1)$$
This yields the condition that
$$\alpha \in Q_{\alpha},$$
where
$$
Q_{\alpha} = \{ \alpha = 2(\gamma^{-1} - 1) | \gamma \in Q_{\gamma} \} .
$$
We also obtain a prime number condition on $w$. This follows from the condition $\alpha = 3(1+w) -2$ in Theorem 3 for $w >1$, which implies,
$$
w = \frac{\alpha - 1}{3}  .
$$
\medskip\medskip

\noindent
This implies that for $w>1$, using the prime number condition in $Q_{\alpha}$ for $\alpha$,
$$
w \in Q_w = \{ w = \frac{2}{3}\gamma^{-1} - 1 | \gamma \in Q_{\gamma} \}  .
$$
that was given in Theorem 1. By Theorem 4, branch regularizability is guaranteed. 
\medskip

\noindent
It is noted that for the case of $w \leq 1$ in Theorem 3, 
$\gamma = 1/3 \in  Q_{\gamma}$ guaranteeing branch regularizability. 
\medskip

\noindent
This proves Theorem 1.
\medskip\medskip

\noindent
{\em Why is the  condition that  $\gamma \in  Q_{\gamma}$ required?}
\medskip

Let $a = F_1(\tau)$ be the singular solution branch $f_1(t)$ in
the scaled time variable $\tau$, where
$F_1 \rightarrow 0$ as $\tau \rightarrow 0^+$, ( $t=0$ is mapped into
$\tau =0$).  We prove below, using the {\em stable
manifold theorem}, what
the exact form of $F_1$ is in terms of powers of $\tau$ which is given by (\ref{eq:Solution}). This yields an expression for $F_1(\tau)$ where the term ${\tau}^{\gamma}$ appears. If $\gamma$ were irrational or of the form $p/q$, $p>0,q>0$ with $q$ even, $p,q$ relativly prime, then ${\tau}^{\gamma}$ would not be defined for $\tau <0$, and $F_1(\tau)$, $\tau \geq 0$, could not be extended to a branch $F_2(\tau)$ for $\tau < 0$. This gives an obstruction to the continuation of a solution $a(\tau)$ through $\tau = 0$ due to the appearance of imaginary time.
\medskip\medskip

\noindent
We now state the result on the transformed flow and the form of the solution $a(\tau)$.
\medskip\medskip\medskip

\noindent
{\bf Theorem 5} \hspace{.1in} $T_M$, together with the transformation of time from $\tau$ to $s$, $d\tau = r ds$, maps (\ref{eq:HamSys}) into the first order system
\begin{equation}
\frac{dr}{ds}  = (\beta + 1) r v,  \hspace{.2in}   \frac{dv}{ds} = \beta(v^2 - 2) - G(r),
\label{eq:TransSys}
\end{equation}
defined on $M$, where  
\begin{equation}
G(r) = \begin{cases} a_1 r^{} + a_2 r^{\frac{2}{3}} + a_3 r^{\frac{\kappa_1}{3}}, &\mbox{if} \hspace{.1in} w < 1,\\
b_1 r^{} + b_2 r^{\frac{2}{3}}, &\mbox{if} \hspace{.1in} w = 1,\\
c_1 r^{\kappa_2 \gamma + 3} + c_2 r^{\kappa_2 \gamma +2} + c_3 r^{\kappa_2 \gamma}, &\mbox{if}\hspace{.1in} w > 1, \end{cases}
\label{G123}
\end{equation}
\medskip

\noindent
The solution for $a = F_1(\tau) \rightarrow 0$ as $\tau \rightarrow 0^+$ can be written 
in explicit form as
\begin{equation}
a = F_1(\tau) = \tau^{\gamma}\Psi(\tau^{\omega_1} , \tau^{\omega_2}) , \hspace{.2in}  \tau > 0, 
\label{eq:Solution}
\end{equation}
$(\omega_1, \omega_2) = (\alpha\gamma/4, \kappa_1/3), (1/3, 1), (\kappa_2\gamma, \gamma)$ for $w<1, w=1, w>1$, respectively, where $\Psi(X,Y)$ is a uniquely determined real analytic function of $X,Y$ for $(X,Y)$ in a sufficiently small neighborhood of $(0,0)$, 
$$
\Psi(0,0) = (\sqrt{2}(\beta + 1))^{\gamma} .
$$
If $\gamma \in Q_\gamma$, then the  unique branch extension of $F_1(\tau)$ to $\tau <0$, where  $|\tau| \ll 1$, is given by
\medskip\medskip

\noindent
$a = F_2(\tau) = F_1(\tau) > 0 $ \hspace{.2in}  if $p$ is even,
\medskip\medskip

\noindent
$a= F_2(\tau) = -F_1(\tau)  > 0 $ \hspace{.2in} if $p$ is odd,
\medskip\medskip

\noindent
where
$F_1(0) = F_2(0) = \Psi(0,0)$. 
\medskip\medskip\medskip
\noindent
\medskip\medskip

\noindent
{\em Proof of Theorem 5}
\medskip

\noindent
The explicit solution $a=a(\tau)$, given by (\ref{eq:Solution}), is obtained by considering (\ref{eq:TransEnergy}) and (\ref{eq:EnergyTerm}), in the original time variable $\tau$.
This implies,
\begin{equation}
\frac{dr}{d\tau} = (1 + \beta)( 2 + 2hr^{\alpha\gamma} + \tilde{G}(r))^{1/2}.
\label{eq:rdot}
\end{equation}
It is noted that $\alpha = 4, \gamma = 1/3$ for $w \leq 1$ and $\alpha = (4 + \kappa_2), \gamma = (1 + \alpha/2)^{-1}$ for $ w > 1$, $\kappa_2 > 0, 0 < \gamma < 1/3$. It is verified that for $w < 1$, the right hand side of 
of (\ref{eq:rdot}) can be written as a analytic function $\tilde{f_1}(X, Y)$, where  $X = r^{\alpha\gamma/4}, Y = r^{\kappa_1/3}$; for $w=1$, it can be written as an analytic function $\tilde{f_2}(X, Y)$, where $X = r^{1/3}, Y = r$; and for  
$w > 1$, it can be written as an analytic function $\tilde{f_3}(X, Y)$, $X = r^{\kappa_2\gamma}, Y = r^{\gamma}$. The functions $\tilde{f_k}(X,Y)$ are real analytic for $(X,Y)$ sufficiently near $(0,0)$ (to insure the square root is well defined) where $\tilde{f_k}(0,0) = \sqrt{2}(1 +\beta)$. More exactly,
\begin{align*}
\tilde{f}_1 &= (1 +\beta)(2 + 2hX^4 + a_1 X^3 + (1/2)a_2 X^2 + a_3(\kappa_1 - 4)^{-1} Y)^{1/2}, \\
\tilde{f}_2 &= (1 +\beta)(2 + 2hX^4 + b_1 Y + b_2 X^2)^{1/2},\\
\tilde{f}_3 &= (1 +\beta)(2 + 2hXY^4 + c_1XY^3 + (1/2)c_2XY^2 + (1/4)X)^{1/2}.\\
\end{align*}
These functions can all be expanded as a convergent Taylor series about (X,Y) = (0,0) with a positive radius of convergence.  Thus, (\ref{eq:rdot}) can be written as
\begin{equation}
\frac{dr}{d\tau} = \mathcal{F}(r^{\omega_1},r^{\omega_2}),
\label{eq:rdot2}
\end{equation}
where $\omega_1(w) > 0, \omega_2(w) > 0$, and $\mathcal{F}(0,0) = \sqrt{2}(1 +\beta)$, $\mathcal{F}(X,Y)$ is real analytic for $(X,Y)$ sufficiently near $(0,0)$. ($(\omega_1, \omega_2) = (\alpha\gamma/4, \kappa_1/3), (1/3, 1), (\kappa_2\gamma, \gamma)$ for $w<1, w=1, w>1$, respectively.) 
We can now apply the stable manifold theorem to (\ref{eq:rdot2}) in an analogous manner that was done in McGehee(1981)(see Lemma 5.7). This implies that there exists an analytic solution,   
\begin{equation}
r(\tau) = \tau \mathcal{G}({\tau}^{\omega_1},{\tau}^{\omega_2}),
\label{eq:rsoln}
\end{equation}   
for $\tau$ in a sufficiently small neighborhood of $0$, $\mathcal{G}(0,0) = \sqrt{2}(1 + \beta)$. 
Setting $a = r^\gamma$, we obtain
$$
a = \tau^{\gamma} \mathcal{G}^{\gamma}({\tau}^{\omega_1},{\tau}^{\omega_2}),
$$
and setting $\Psi = \mathcal{G}^{\gamma}$ yields the solution in Theorem 5. It is necessary to prove that $\omega_1, \omega_2$ insure that $\tau^{\omega_k}, k=1,2,$ is not imaginary for $\tau < 0$. This follows immediately for $w = 1$, since $\omega_1 = 1/3, \omega_2 = 1$. For $w < 1$, $\omega_1 = \alpha \gamma/4 = \gamma$ and $\omega_2 = \kappa_1 /3 = 1- w$. But $w = (2/3)(p/q) -1$ so that $\omega_2 = 2[1 - (1/3)(p/q)].$ Finally, for $w > 1$, $\omega_2 = \gamma$ and $\omega_1 = \kappa_2 \gamma = 3(w-1) \gamma = 3(2[(1/3)(p/q) - 1])\gamma = 6[(1/3)(p/q) - 1]\gamma.$ 
\medskip

\noindent 
The system (\ref{eq:TransSys}) of differential equations is obtained by transforming (\ref{eq:HamSys}) using (\ref{eq:Trans}), together with the transformation of time from $\tau$ to $s$, which is verified after some algebraic simplification. 
\medskip

\noindent
The concludes the proof of Theorem 5.
\medskip\medskip\medskip

\noindent
{\em Remark} \hspace{.1in}  It is noted that the key result in Theorem 5 is Equation (\ref{eq:Solution}) for $\tau \geq 0$ as 
$\tau \rightarrow 0^+$ and its continuation, $F_2(\tau)$, for $\tau \leq 0$ as  $\tau \rightarrow 0^-$, in the original physical coordinates $a,\tau$. The
flow in the regularized coordinates $r,v,s$ given by Equation (\ref{eq:TransSys}), described below, is not required to conclude the extension of the solution. However, (\ref{eq:TransSys}) is necessary to understand the nature of the flow near $a =0$. 
\medskip
\medskip

Thus, by Theorem 5, $a(\tau) = F_1(\tau) \rightarrow 0$ as $\tau \rightarrow 0^+$, and at $\tau = 0$, $F_1(0) = 0$, then as $\tau$ passes to negative values of
$\tau$, $a(\tau) = F_2(\tau)$  increases.  This motion of $a$ provides a universe bounce through $a=0$ through a unique branch extension. This motion as a function of
$\tau$ can be done as function of $t$ through the constant scaling (\ref{eq:TimeScale}).  By Theorem 4, the
condition  $\gamma \in  Q_{\gamma}$ is required. The bounce is nondifferentiable at $\tau =0$, or $t =0$,  which is seen by differentiation of $F_1$ with 
respect to $\tau$.  This introduces the term $\tau^{\gamma - 1}$ that multiplies $F_1$. Since $\gamma < 1$, then $\tau = 0$ is undefined.
\medskip\medskip

\noindent
{\em Geometry of flow in branch regularized coordinates}
\medskip\medskip

\noindent
{\bf Lemma 1} \hspace{.1in} $N$ is an invariant manifold for the flow of (\ref{eq:TransSys}). As a trajectory approaches $N$ as $\tau \rightarrow 0^+$, then this implies $s \rightarrow  \infty$. 
\medskip\medskip

\noindent
This is proven from the definition of $N$, which implies $r =0$ and $v=\pm \sqrt{2}$. These are rest points for (\ref{eq:TransSys}). Thus, $N$ is an invariant manifold set for (\ref{eq:TransSys}). This implies that in the time variable $s$, $s \rightarrow \infty$ as $\tau \rightarrow 0^+$. 
\medskip

It is noted that the vector field defined by the system of differential equations (\ref{eq:TransSys}) is not differentiable at $r=0$, with respect to $r$, due to the term $G(r)$. This is seen by noting that $dG/dr$ is singular at $r=0$. It is, however, locally Lipschitz continuous in a neighborhood of $r=0$, guaranteeing uniqueness.  It is real analytic with respect to $v$ at $v= \pm\sqrt{2}$ and $r=0$. 
\medskip\medskip

It is remarked that this branch regularization can be viewed as a minimal type regularization where the flow in the regularized coordinates can be smoothed to the extent where the
collision set results in rest points. This yields unique branch extensions in physical space which are continuous but not differentiable. 
In other problems, eg the Newtonian inverse square central force
problem, collision can be regularized so that in the regularized coordinates, the flow can be extended through the transformed collision set in finite time, where
the flow is smooth in a full neighborhood of collision. These are 'global' regularizations, where, moreover, the flow is smooth on the entire energy manifold. Examples of these are mentioned in the Introduction, eg the Levi-Civita regularization or the geodesic equivalent flow maps (Levi-Civita 1920; Moser 1970; Belbruno 1977) (see also Belbruno 2004). These stronger regularizations also imply that in the original coordinates for the Newtonian inverse square central force law, each
collision solution can be extended through collision in a bounce as in the weaker branch regularization. The difference is that the global regularization yields
a smoother behavior of the flow near collision than in the branch regularization.

It is important to mention that another type of regularization of interest is 'block regularization'. In this case the flow of the differential equations is viewed in a compact neighborhood of the singular set, called an isolating block.  In one version of this, due to C. Conley and R. Easton (Conley and Easton 1971), the singular set is an invariant set where the vector field is defined. In another variation of this, due to R. Easton (Easton 1971) the vector field need not be defined on the singular set. In either case, the flow is studied near the singular set within the isolating block. The flow is called {\em block regularizable}, in a deleted neighborhood of the singular set, if it can be shown to be diffeomorphic to the trivial parallel flow. It is proven in McGehee(1981), in the case of a pure power law, $a^{-\alpha}$, given by (\ref{eq:Pot}) with $f(a) = 0$, that the flow is block regularizable if and only if $\beta = 1- n^{-1}, n=1,2,3,... $.   This is done by showing an equivalence with the flow resulting from applying a Levi-Civita type regularizing transformation. A similar result may true in our case when $f(a)$ is included. However, in this paper, $\beta \geq 2.$ This result seems to suggest that when $f(a)$ is included, the flow may not be block regularizable, although this is not proven. 

Block regularization does not require looking at the flow at the singularity, but near it. If it is block regularizable, then this implies a degree of uniformity of the flow near the singularity. One could make the rough analogy that the flow could be combed.  Physically, this would imply that the dynamics of the system near the big bang, for example, would be fairly well behaved. If it were not block regularizable, then the behavior of the flow near the big bang would not be uniform in the sense that it would not be equivalent to a trivial parallel flow.  

It is noted that branch regularization does not imply block regularization, and although block regularization does imply branch regularization in the case of a pure power law, as is proven in McGehee(1981), it is not necessarily the case in general that block regularization implies branch regularization.
\medskip
\medskip
\medskip

\section{{\bf Relevance of Results }}
\label{sec:3}
\medskip\medskip

The big bang singularity is not understood in reality. The big bang is thought to have occurred approximately 13.75 billion years ago. The initial rapid expansion of the universe after this state, modeled by the theory of inflation, is estimated to have occurred approximately within $10^{-36}$ and $10^{-32}$ seconds, with an expansion factor of minimally $10^{78}$ in volume. The initial temperature near this period is estimated to have been $10^{32}$ degrees Kelvin, beyond our ability to experimentally reach in particle accelerators.  There are a number of cosmological theories that try to understand the big bang. In this paper we have taken a purely mathematical approach, using a Friedmann model which is generally used to understand the expansion and evolution of the universe, and not the big bang singularity itself. We have applied a McGehee regularization transformation to this model, to try and understand the dynamics in a neighborhood of the big bang singularity, given by $a=0$. This mathematical analysis does provide information on the flow of the differential equations we are using in this neighborhood, derived from the Friedmann model, yielding conditions on which solutions can be mathematically extended through the singularity. The existence of such solutions  is not intended to be applicable to reality. Our analysis does not attempt to explain any of the physics or nature of this state.   

It is instructive to make a very brief comparison of our methodology and results to other cosmological models. For a comparison, we will use the Friedmann equation given by (\ref{eq:H2}), which results from general relativity. We will also use the equation of state $w$, since that plays a key role in our results, and also for the sake of simplicity. The parameter $w$ was defined in the introduction.  For a static universe, $w = -1/3$. When the universe expands, $ w < -1/3$. In Theorem 1, $w = 1$ is a transition case between extensions of solutions through $a = 0$ that can be done, in general, for $w \leq 1$ and for $ w > 1$ where it is necessary that $w \in Q_w$. As is described in Erickson et al.(2004), for $w < 1$, the anisotropy and curvature terms in the Einstein field equations grow rapidly and become dominant as $a \rightarrow 0$, that can cause chaotic behavior that may destabilize the contraction process. However, if $w > 1$, these terms remain negligible as compared to the energy density, $\rho$. In this sense, $w=1$ represents a transition. This is interesting to compare to the role $w=1$ plays as being a transition for conditions required for branch regularization. 

The cosmological theories we will be comparing too model the evolution of the universe using a Freidmann equation, however, their modeling of the big bang singularity are all quite different.  It is important to note that there are many other factors that can be used in making such comparisons between theories, however, since we are not attempting to understand the big bang, the Friedmann equation for the evolutionary behavior plays a key role. Further analysis beyond this comparison is beyond the scope of this paper. There are many cosmology theories, and we are considering a subset for comparison.    

In Gott and Li(1998),  J. R. Gott and L. X. Lin discuss the possibility that our universe could expand from $a=0$, then stop expanding and eventually contract into a 'big crunch' again to $a=0$. This could repeat cyclically. In the contraction phase immediately immediately before $a=0$, if $w > 1$, then by our Friedmann model, Theorem 1 would imply that for an extension through $a=0$, it is necessary that $ w \in Q_w$, a set of measure zero. However, if, as $a \rightarrow 0$, $w < 1$, then branch extension can be done in general. However, in this case, chaotic behavior can occur.   

P. Steinhardt, N. Turok, have put forward the ekpyrotic theory of the big bang (Steinhardt et al. 2001). It is a cyclic universe theory that uses string theory. In this theory, the big bang is hypothesized to have occured cyclically by the collision of branes, which are subspaces derived from string theory. The resulting expansion and contraction is modeled using a Friedmann equation that is given by (\ref{eq:H2}).  In this theory, $w > 1$ and $w \rightarrow 1$ as $a \rightarrow 0$ (see Lehners 2009; Steinhardt and Turok 2005). This yields an interesting interpretation of results when compared to our methodology: If the convergence of $w$ to $1$ is within the tolerances required by physical considerations, including quantum fluctuations,  yielding $w =1$ in a neighborhood of $a = 0$, which is claimed, then an extension through $a=0$ is automatic by Theorem 1. However, if $w > 1$ is maintained for $a > 0$ in the limit, no matter how small, then $w$ would need to satisfy relative prime number values on the set $Q_w$ prior to $a = 0$ to be consistent with Theorem 1, placing a restriction that $w \in Q_w$.  This condition, however, may be meaningless, since when $a$ goes below the Planck scale, the laws of physics are no longer well defined.  It is important to note that although there are issues for the ekpyrotic/cyclic models for $w >1$ in our analysis, quantum corrections, for any value of $w$, still allow an effective leading order description of the evolution of the universe in terms of the scale factor, $a$,  governed by the Friedmann equation.

There are other cosmological theories that use string theory. 
For example, another is string gas cosmology (Brandenberger 2004) put forth by R. Brandenberger and C. Vafa.  This is a cosmological theory that incorporates a gas of strings within the four-dimensional space-time of the early stages of the inflation of the universe. The additional dimensions given by strings, offers a way to explain aspects of the universe in a way that is different than just looking at a pure inflationary model. The reduction of the evolution of the universe based on a string gas theory to a Friedmann equation was carried out by A. Kamenshchik and I. Khalatnikov (see Kamenshchik and Khalatnikov 2011) by viewing a string gas as a perfect fluid with $w = -1/3$. The Friedmann equation they obtain is different from (\ref{eq:H2}), so that we would have to repeat our analysis with this equation to make conclusions on branch regularizability. Other theories using string theory include, brane cosmology and string landscape, which are beyond the scope of this paper.

Another theory of the big bang we consider is by J. Hartle and S. Hawking (Hartle and Hawking 1983). Their theory models the big bang using a quantum mechanical framework, where our universe is viewed as a particle and the big bang is modeled as a wave function. The evolution of the universe after the big bang uses general relativity. The Hartle-Hawking theory assumes that time does not exist prior to the big bang singularity. This condition is inconsistent with the methodology of our results since we require a continuous time flow through the big bang. However, another theory of interest of the big bang using quantum mechanical ideas and quantized gravity, is called loop quantum cosmology (Thiemann 2007). It is based on loop quantum gravity, which is a quantization of general relativity, developed by A. Ashtekar. This theory gives rise to a cyclical universe
theory where time can be continued though the big bang. This approach can be reduced to a modified Friedmann equation (Singh 2005). 
It is shown in Singh(2005) that $w \in [-1, 1]$. Within our methodology, these values of $w$ this would automatically allow an extension of solutions through the big bang singularity. However, the Friedmann equation used in this theory is a modification of (\ref{eq:H2}), and we would have to repeat our analysis for this equation to make definitive conclusions. 

It is noted that Theorem 1 implies a 'unique' branch extension from a previous universe to ours. As noted in the previous section, uniqueness results from Libschitz continuity of (\ref{eq:TransSys}). A modification of the original Friedmann equation could result in nonuniqueness of solutions at $a=0$ for the transformed differential equations.  Also, we have fixed a number of variables as $a \rightarrow 0$, such as $w$, and the curvature $K$, among others. Although fixing these parameters seems consistent with numerical simulations in Garfinkle et al.(2008), it would be interesting to study the approach taken in this paper if they were allowed to vary. This is a topic for further study.

\medskip
\medskip
\medskip

\noindent
{\bf Acknowledgements}
\medskip

\noindent
I am grateful for discussions with David Spergel, Paul Steinhardt, and Frans Pretorius.  This work was supported by NASA/AISR grant NNX09AK61G.

\medskip
\noindent

\end{document}